\documentclass[epjCONF]{svjour}
\usepackage{graphics}
\usepackage[varg]{txfonts} 
\usepackage[latin1]{inputenc}
\usepackage{natbib}

\newcommand{\apj}{ApJ}
\newcommand{\aap}{A\&A}
\newcommand{\mnras}{MNRAS}
\newcommand{\nat}{Nature}
\newcommand{\apjl}{ApJL}
\newcommand{\apjs}{ApJS}
\newcommand{\aj}{AJ}

\session-title{CoRoT Symposium 3, Kepler KASC-7 joint meeting}
\begin{document}
\title{Stellar magnetic activity -- Star-Planet Interactions}
\author{Poppenhaeger, K.\inst{1}\fnmsep\inst{2}\fnmsep\thanks{\email{kpoppenhaeger@cfa.harvard.edu}} }
\institute{Harvard-Smithsonian Center for Astrophysics, 60 Garden Street, Cambrigde, MA 02138, USA \and NASA Sagan Fellow}
\abstract{
Stellar magnetic activity is an important factor in the formation and evolution of exoplanets. Magnetic phenomena like stellar flares, coronal mass ejections, and high-energy emission affect the exoplanetary atmosphere and its mass loss over time. One major question is whether the magnetic evolution of exoplanet host stars is the same as for stars without planets; tidal and magnetic interactions of a star and its close-in planets may play a role in this. Stellar magnetic activity also shapes our ability to detect exoplanets with different methods in the first place, and therefore we need to understand it properly to derive an accurate estimate of the existing exoplanet population. I will review recent theoretical and observational results, as well as outline some avenues for future progress.
}
\maketitle
\section{Introduction}
\label{intro}

Stellar magnetic activity is an ubiquitous phenomenon in cool stars. These stars operate a magnetic dynamo that is fueled by stellar rotation and produces highly structured magnetic fields; in the case of stars with a radiative core and a convective outer envelope (spectral type mid-F to early-M), this is an $\alpha\Omega$ dynamo, while fully convective stars (mid-M and later) operate a different kind of dynamo, possibly a turbulent or $\alpha^2$ dynamo. These magnetic fields manifest themselves observationally in a variety of phenomena. Prominent examples are: starspots, which are a manifestation of magnetic field lines piercing the stellar photosphere and obstructing the convective welling up of hot plasma, causing these spots to cool and become darker than the surrounding photosphere; heating of the stellar chromosphere and the corona, causing emission in chromospheric lines, the UV, and at X-ray energies; as well as impulsive flares, visible all across the spectrum from the radio to X-rays, caused by reconnection of magnetic field lines.

Cool stars shed an ionized stellar wind, which couples to the stellar magnetic field on its journey away from the star and thus carries away angular momentum. The star therefore spins down over time. It is an active field of study in astrophysics to understand precisely how stellar rotation and magnetic activity evolve over time. A proper understanding of this evolution requires measurements of the rotation and activity itself, and also a reliable measurement of the stellar ages to fix the time coordinate. For the first part, one typically uses rotational periods or projected rotational velocities \citep{Skumanich1972}, X-ray emission \citep[for example]{Preibisch2005}, chromospheric line emission in Ca~II H and K or H~$\alpha$ \citep[for example]{Mamajek2008}, or magnetic field strengths measured through Zeeman splitting \citep{Vidotto2014}. For the second part (measuring ages), one can use the presence of protoplanetary disks or lithium in the stellar photosphere to identify young star. It is also possible to determine a young stellar age by association if a star belongs to a young stellar cluster. Ages from a few Myr to several 100 Myr are measurable like that, and yield insights into the rotational periods \citep[for example]{Bouvier1986, Carpenter2001, Rebull2014, Cody2014} and X-ray activity \citep[for example]{Stelzer2001, Preibisch2005} with which stars enter the main sequence. For older stars, the situation becomes more difficult. Old stellar clusters are hard to identify and observe because they are either very spread out or far away, although some progress has been made recently using \textit{Kepler} data \citep[for example]{Meibom2011_1Gyr}. More fuzzy age estimates are possible through moving groups, or association with the old disk or galactic halo population. Recent progress has also been made in estimating the ages of individual old stars through asteroseismology using high-precision photometric data from \textit{Kepler} and \textit{CoRoT} \citep{SilvaAguirre2013, Chaplin2014}. 

Stellar magnetic activity matters to exoplanets on several axes: activity and rotation of stars provides a possibility to estimate ages of planetary systems; the stellar high-energy emission influences the planetary atmospheres and can lead to atmospheric evaporation; and there are hints that in some exoplanet systems the planets can even influence the stellar activity. In the following sections I will review observational results in these three areas.

\section{Star-planet interactions: planets influencing their host stars}

There is a class of stars that does not follow the typical evolution of rotation and activity of cool stars, namely close stellar binaries. In these systems the two stars orbit each other with a few days period, and they rotationally synchronize over time through tidal interaction, i.e.\ the stellar rotation period matches the orbital period of the binary. This keeps the stars spinning fast even at an old age, and their activity therefore stays at a high level. Observations of such systems have found observational signatures of direct magnetic interaction between the two stars as well \citep{Siarkowski1996}. This is the template for star-planet interactions: one thinks of a star with a Hot Jupiter as a binary with a very small mass ratio, for which one may expect to see signatures of tidal and magnetic interaction at some level \citep{Cuntz2000, Lanza2008}. Detecting such interactions and determining their strengths can yield insights into the stellar and planetary magnetic fields involved \citep{Lanza2009}.

There have been extensive observational searches for effects of tidal or magnetic star-planet interaction. The arising picture shows some very interesting hints for such interactions, but also non-detections in systems where one expects to see such signatures. Some facets of the picture are summarized here.

Initial repeated observations of Hot Jupiter host stars found chromospheric activity peaks phased with the orbital period of the planet (not the rotational period of the host star) for two of the targeted systems, namely HD 179949 and $\upsilon$ Andromedae \citep{Shkolnik2005}. However, in later observations only periodic signatures compatible with stellar rotation were detected \citep{Shkolnik2008, PoppenhaegerLenz2011, Scandariato2013}. In observations of the Hot Jupiter host star HD 189733, repeated stellar flares were reported after the eclipse of the planet, possible indicating a stellar hot spot or a long connecting magnetic loop being dragged around by the planet \citep{Pillitteri2010, Pillitteri2011, Pillitteri2014}. Tidal interaction has been suspected to cause the matching of the orbital period of the Hot Jupiter $\tau$ Boo b and the rotational period of its host star \citep{Barnes2001}, and increased stellar variability has been suspected to be caused by the planet \citep{Walker2008}. On the other hand, a system with a very close and massive Hot Jupiter, WASP-18, did not display any chromspheric or coronal activity enhancements \citep{Miller2012, Pillitteri2014WASP18}. 

The community has also tried to tackle this problem by observing large samples of planet-hosting stars to test for trends related to star-planet interactions. In observations of coronal activity for a large sample of planet-hosting stars, Hot Jupiter-hosting stars were found to be systematically more X-ray active than stars with small, far away planets \citep{Kashyap2008, Scharf2010}. Later volume-complete studies recovered the same trend; however, the trend could be traced back to selection effects related to the planet detection efficiency \citep{Poppenhaeger2010, Poppenhaeger2011}. Especially for nearby exoplanets, where the radial velocity (RV) method is historically the dominant detection mechanism, it is true that stellar activity masks the RV signal of the planets. This means that we are biased to find only the massive, close-in planets around active stars, while the small, far away planets go undetected. A par\-ti\-cu\-lar\-ly enlightening example is the super-earth CoRoT-7 \citep{Alonso2008} that orbits an X-ray active star \citep{Poppenhaeger2012}; it was first detected through a transit light curve, and even though the presence of this planetary candidate was already known, the detection of its RV signature proved to be extremely challenging \citep{Queloz2009, Hatzes2010, Ferraz-Mello2011}. For inactive stars, one is generally able to detect both the Hot Jupiter-type planets and the smaller planets in wider orbits. This difference in detection efficiency for active and inactive host stars skews the known pool of exoplanets significantly. 

While studies of stellar samples indicate that star-planet interactions are not a strong factor through the whole population of planet-hosting stars, the observations of individual systems indicate that at least for some systems with particularly close-in planets star-planet interactions may be a relevant factor.

\section{Star-planet interactions: the reliability of rotation/activity-derived ages}

\begin{figure}
\begin{center}
\resizebox{0.75\columnwidth}{!}{%
\includegraphics{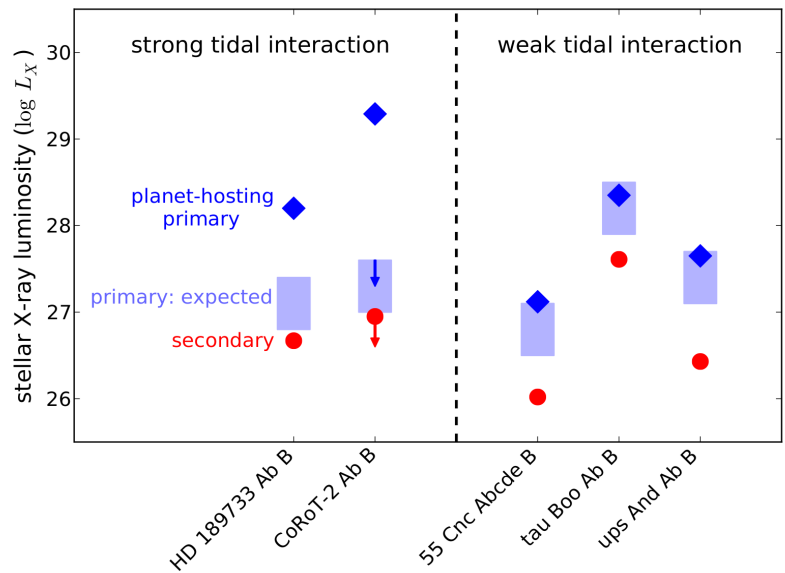}}
\caption{Stars that interact tidally with their Hot Jupiters may display excess activity and rotation when compared to stellar companions with the same age (left part of this figure). The secondary star (without known exoplanets) acts as a negative control against which the activity and rotation of the planet-hosting star can be comnpared \citep{Poppenhaeger2014}.}
\label{fig:0} 
\end{center}
\end{figure}

An important point for systems with very close-in exoplanets is whether there is a significant tidal interaction that may have altered the rotational evolution of the host stars. Such an interaction would be mediated through tidal bulges raised by the planet on the stellar surface, and angular momentum transfer from the planetary orbit to the stellar spin might occur. This is thought to be much more efficient for cool stars with outer convection zones than for hot stars which possess an outer radiative layer \citep{Zahn2008}.

Intriguing data has been presented showing that the misalignment between the stellar spin axis and the planetary orbital plane is generally small for cool stars with Hot Jupiters, but the full range of possible misalignments is found for hot stars without significant outer convection zones that host Hot Jupiters \citep{Winn2010, Albrecht2012}. This has been interpreted as an effect of tidal alignment over time.

Studies of several indidvual systems have shown that some Hot Jupiter host stars are rotating faster than expected \citep{Pont2009tidal, Brown2011, Husnoo2012}. It is possible to control for selection effects due to stellar activity by observing wide stellar binaries in which one of the stars hosts an exoplanet, while the other star does not possess any known exoplanets. In such a case, the second star acts as a negative control, and (after adjusting for possibly different spectral types of the two stars) an excess rotation or activity of the planet-hosting star is likely caused by a planetary influence. \cite{Poppenhaeger2014} studied the activity and rottaion in five such systems, out of which two were expected to experience strong tidal interaction between planet and host star. Indeed, those two systems displayed excess activity and rotation of the planet-hosting stars, while the systems with expected weak tidal interaction displayed no such discrepancies (see Figure~\ref{fig:0}). 

It is therefore possible that the typical age-rotation-activity relationships of cool stars are not applicable for stars with strong tidal interaction with their planets. Age measurements through asteroseismology will be important to provide an independent age diagnostic to be calibrated against stellar rotation and activity \citep{SilvaAguirre2013, Chaplin2014, McQuillan2014}.

\section{Star-planet interactions: stars influencing their planets}

\begin{figure}
\begin{center}
\resizebox{0.85\columnwidth}{!}{%
\includegraphics{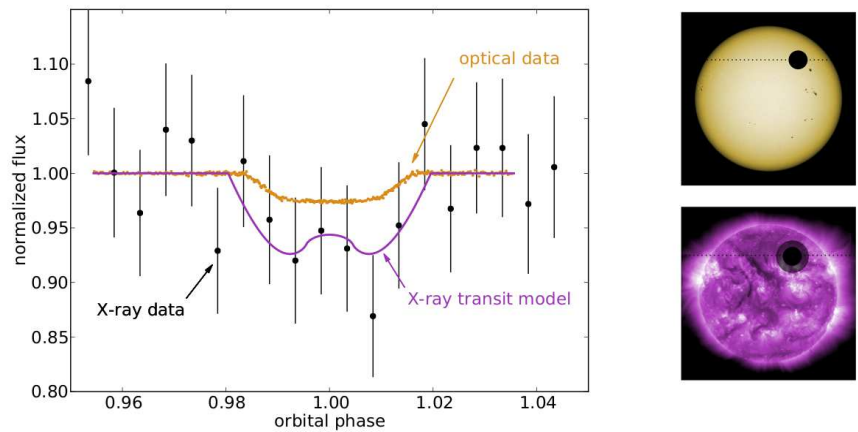}}
\caption{Exoplanet transit observations at short wavelengths (here shown for the transiting Hot Jupiter HD~189733~b in the  X-ray band) reveal the extent of outer layers of exoplanetary atmospheres, as short-wavelentgh photons are absorbed high up in the atmosphere \citep{Poppenhaeger2013}.}
\label{fig:1}  
\end{center}
\end{figure}

A high level of magnetic activity of the host star will influence its exoplanets. For this, it does not matter if the high activity is a product of stellar youth or of some planetary influence on the host star itself. A variety of effects have been investigated, including the effect of X-ray and UV (XUV) radiation on exoplanetary atmospheres \citep{Lammer2003, Lecavelier2004, Tian2005, Erkaev2007, Murray-Clay2009}, the sudden effects of stellar flares \citep{Segura2010}, atmospheric stripping by stellar winds or coronal mass ejections \citep{Khodachenko2007, Lammer2007}, and theoretical studies of magnetic effects like Joule heating of atmospheres \citep{Cohen2014}. 

A major area of interest is the atmospheric evaporation of exoplanetary atmospheres, which is thought to be primarily driven by X-ray and extreme UV photons from the host star \citep{Lecavelier2004}. These photons are absorbed in the higher layers of the planetary atmosphere, where they cause strong heating and can lift material out of the gravitational well of the planet. The extent of these exospheres can be measured using transit obervations at UV and X-ray wavelengths, and indeed X-ray and UV radii that are 40\%-70\% larger than the optical radius have been observed for Hot Jupiters \citep{Vidal-Madjar2003, Lecavelier2010, Fossati2010, Haswell2012, Poppenhaeger2013, Bourrier2013}; see Figure~\ref{fig:1} for X-ray transit observations fo the Hot Jupiter HD 189733 b. 

For smaller exoplanets like mini-neptunes and super-earths such observations are currently very challenging due to the required high signal to noise. However, theoretical models of exoplanetary mass loss can be employed to yield a mass loss estimate based on the stellar X-ray and EUV luminosity \citep{Lecavelier2004, Lopez2013}. Recent observations of the super-earth host star GJ 1214 have shown that the XUV irradiation of the planet GJ 1214 b is at least five times higher than for the Hot Jupiter HD 209458 b \citep{Lalitha2014}, which is known to be actively evaporating \citep{Vidal-Madjar2003}; with very sensitive observations it will therefore be possible to observe the mass loss of small planets as well, as the energy budget supplied at the top of the atmosphere should be sufficient for evaporation.

Another topic that is being explored is the response of planetary atmospheres to impulsive stellar flares, which can cause a sudden dramatic increase in high-energy irradiation. This is of particular interest for planets orbiting M dwarfs, because the habitable zone is very close to the host star. M dwarfs stay magnetically active for a long time \citep{West2008} and are known to produce strong variability in the UV and in X-rays \citep{Guedel2004, Fuhrmeister2011, France2013}. Theoretical models have shown that such flares have the potential to alter the chemistry of exoplanet atmospheres \citep{Miguel2014arXiv} and that it is possible for significant UV fluxes to reach the planetary surface \citep{Segura2010}.

\section{Summary}

Stellar magnetic activity is an important factor when trying to understand exoplanetary systems. The activity-induced high-energy emission of the host star can influence exoplanetary atmospheres through stellar flares and can cause atmospheric evaporation. On the other hand, it is possible that massive, close-in planets can influence the stellar rotation and activity through star-planet interactions. The current observational landscape shows some intriguing data for individual systems, but a consistent picture about when and where to expect star-planet interactions to occur still has to arise. Progress will be made through in-depth analyses of stellar ages and the expected rotation and activity properties at a given point in the life of a star, and through exoplanet detection missions that will focus on nearby systems, so that the host stars of these systems can be studied in detail.

%
%

\end{document}